\documentclass[preprint]{elsarticle}

\usepackage{amsfonts}
\usepackage{amsmath}
\usepackage{amssymb}
\usepackage{amsthm}
\usepackage{graphicx}
\usepackage{subfigure}
\usepackage{mathdefs}
\usepackage{mathrsfs}
\usepackage{lineno}

\usepackage[noend]{algpseudocode}
\usepackage{algorithm}
\usepackage{placeins}
\usepackage{bm}
\usepackage{todonotes}
\usepackage{enumitem}
\usepackage{booktabs}
\usepackage{verbatimbox}

\usepackage{dsfont}
\usepackage{colortbl}
\definecolor{lg}{gray}{0.9}

\newtheorem{definition}{Definition}

\newtheorem{counter_ex}{Counterexample}

\newcommand{\eqnref}[1]{(\ref{eq:#1})}
\newcommand{\secref}[1]{\S\ref{sec:#1}}

\newcommand{\figref}[1]{Fig.~\ref{fig:#1}}
\newcommand{\tabref}[1]{Table~\ref{tab:#1}}
\newcommand{\algoref}[1]{Alg.~\ref{alg:#1}}

\newcommand{\linref}[1]{Line~\ref{lin:#1}}
\newcommand{\ceref}[1]{C.E.~\ref{ce:#1}}

\usepackage{hyperref}
\hypersetup{colorlinks=false}
\hypersetup{colorlinks,citecolor=black,filecolor=black,linkcolor=black,urlcolor=black}

\journal{Information Processing Letters}
\begin{document}
\begin{frontmatter}

\title{Common greedy wiring and rewiring heuristics do not guarantee maximum assortative graphs of given degree}
\tnotetext[t1]{This work was supported by the National Science Foundation, Grant $\#$IIS-1250786.}

\author{Jonathan Stokes}
\author{Steven Weber}
\address{Department of Electrical and Computer Engineering}
\address{Drexel University, Philadelphia, PA, USA}
\begin{abstract}
We examine two greedy heuristics --- wiring and rewiring --- for constructing maximum assortative graphs over all simple connected graphs with a target degree sequence.  Counterexamples show that natural greedy rewiring heuristics do not necessarily return a maximum assortative graph, even though it is known that the meta-graph of all simple connected graphs with given degree is connected under rewiring.  Counterexamples show an elegant greedy graph wiring heuristic from the literature may fail to achieve the target degree sequence or may fail to wire a maximally assortative graph.
\end{abstract}
\end{frontmatter}

\section{Introduction}
\label{sec:intro}

\subsection{Motivation}
The assortativity of a graph (Newman \cite{New2002}) is the correlation of the degrees of the endpoints of a randomly selected edge. High degree vertices tend to be connected to high (low) degree vertices in positively (negatively) assortative graphs.  

One (of many) practical implications of assortativity is in graph search, e.g., searching a (often large order) graph for (one or all) vertices of maximum (or at least large) degree \cite{AvrLit2014}: the work presented in \cite{StoWeb2016b,StoWeb2016c} has studied the performance impact of assortativity on search heuristics such as sampling and random walks.  Finding such vertices in large graphs has diverse applications, including viral marketing in social networks and network robustness analysis \cite{KemKle2003,CohEre2001}, among numerous others.    

The motivation for this paper is the problem of identifying a collection of graphs, all from the class of graphs with a given degree sequence, with the assortativity of the graphs in the collection varying from the minimum to the maximum possible within that class. The performance impact of the assortativity on the search heuristic may be studied by running the heuristic on all graphs in the collection. Given this objective, the first step is to identify graphs with extremal assortativity within the class. This paper examines two greedy heuristics for finding maximum assortative graphs within a class: graph rewiring and wiring.  

\subsection{Related Work}
There is an extensive literature on extremization of assortativity over different graph classes; this section briefly covers the most pertinent points of this literature, focusing on the distinctions between the work presented in this paper and the prior work.

{\em Assortativity.} Newman \cite{New2002} introduced (graph) assortativity which is denoted $\alpha \in [-1,+1]$. Van Meighem  \cite{MieWan2010} showed perfect assortativity ($\alpha=1$) is only possible in regular graphs, while any complete bipartite graph $K_{m,n}$ ($m \neq n$) is perfectly disassortative ($\alpha=-1$). There is a large literature on network degree correlations and assortativity (e.g., \cite{OrsDan2015}), and on graphs with extremal assortativity within a class (e.g., \cite{KinKun2016}).

{\em Joint Degree Matrix (JDM).} The generation of random graphs with a particular JDM (also called a 2K-series) has been the subject of a number of recent papers. Stanton \cite{StaPin2012} and Orsini \cite{OrsDan2015} have proposed random edge rewiring as a method of sampling graphs with a given JDM, while Gjorka \cite{GjoTil2015} has introduced a random wiring method for constructing these graphs.  However, there is no means known to us by which JDMs may be efficiently enumerated, and therefore there is no easy means to maximize assortativity, which is a statistic of the JDM, short of enumerating all (in our case, simple and connected) graphs with a given degree sequence.

{\em Rewiring.} The meta-graph for a degree sequence, with a vertex for each connected simple graph with that degree sequence and an edge connecting graphs related by rewiring a pair of edges, was studied by Taylor \cite{Tay1981}; in particular, he showed this meta-graph to be connected (Thm. 3.3) extending an earlier result by Rysler for simple graphs \cite{Rys1957}. This fact is used in \secref{rewiring}. 

Following Rysler's work, rewiring heuristics for sampling graphs with a particular degree sequence (e.g., \cite{KanTet1999}, \cite{MasSne2002}, \cite{OrsDan2015}) have been introduced. Rewiring heuristics have also been proposed by Newman \cite{New2003}, Xuli-Burnet \cite{XulSok2005}, Van Meighem \cite{MieWan2010}, and Winterbach \cite{Win2012} along others for changing a graph's assortativity. The first three of these algorithms, being purely stochastic, cannot efficiently maximize assortativity. Winterbach's algorithm uses a guided rewiring technique to maximize assortativity. However, this technique does not maintain graph connectivity, as its rewirings are a subset of those explored by rewiring heuristic $A$ (see \secref{grh}), and therefore Winterbach's algorithm does not necessarily maximize assortativity.

{\em Wiring.} Li and Alderson \cite{LiAld2005} introduced a greedy wiring heuristic for constructing a graph with maximum assortativity over the set of simple connected graphs with a target degree sequence. Kincaid \cite{KinKun2016} argues wiring a minimally or maximally assortative connected simple graph is NP-hard and proposes a heuristic which is shown numerically to perform near optimally in minimizing graph assortativity. Winterbach \cite{Win2012}, Zhou \cite{ZhoXu2008}, and Meghanathan \cite{Meg2016} have also proposed methods unconstrained by graph connectivity of wiring maximally assortative graphs. This paper examines Li's heuristic further in \secref{wiring}. 

{\em Graph enumeration and generation.} The results in this paper where achieved using {\sf geng}, a tool in the {\sf nauty} package created by McKay \cite{McKPip2014}, to generate all simple connected graphs of a given order. 

\subsection{Notation}
Let $a \equiv b$ denote equal by definition. Let $[n]^+$ denote $\{1,\ldots,n\}$ for $n \in \Nbb$. A graph of order $n$ is denoted $G = (\Vmc,\Emc)$, with vertices $\Vmc = [n]^+$ and edges $\Emc$; size is denoted by $m = |\Emc|$. A directed edge between vertices $i$ and $j$ is denoted $(ij)$, and an undirected edge is denoted $ij$ or $\{ij\}$.\footnote{Except in \secref{wiring} where \algoref{s_max}'s undirected pedges are listed as an ordered pair.} Let $d_i$ denote the degree of vertex $i$, $\dbf = (d_i, i \in \Vmc)$ denote a degree sequence, and $\dbf_G = (d_i, i \in \Vmc)$ the degree sequence for graph $G$.  Additionally, let $\mathrm{Uni}(\Vmc)$ denote the uniform distribution over vertex set $\Vmc$, $\mathrm{Var}(d_{\wsf})$ be the variance of the degree of a randomly selected vertex $\wsf \sim \mathrm{Uni}(\Vmc)$, and $\mathrm{Corr}(d_{\usf},d_{\vsf})$ be the correlation between the degrees of random vertices $\usf$ and $\vsf$.
 
The collection of distinct unlabeled undirected simple connected graphs of order $n \in \Nbb$ is denoted $\Wmc^{(n)}$. Let $\Dscr^{(n)} \equiv \bigcup_{G \in \Wmc^{(n)}} \dbf_G$ be the collection of degree sequences found in graph collection $\Wmc^{(n)}$, and let $\Wmc^{(n)}_{\dbf} \equiv \{ G \in \Wmc^{(n)}| \dbf_G = \dbf\}$ be the graphs in $\Wmc^{(n)}$ with degree sequence $\dbf$, henceforth referred to as the {\em degree class} $\dbf$. It follows that $(\Wmc^{(n)}_{\dbf}, \dbf \in \Dscr^{(n)})$ is the partition of $\Wmc^{(n)}$ by the degree sequence $\dbf$.

The $S$-metric and assortativity, for $G = (\Vmc,\Emc) \in \Wmc^{(n)}$, are defined below.
\begin{definition}
\label{ch6_assort}
The $S$-metric \cite{LiAld2005} is, 
\begin{equation}
s(G) \equiv \sum_{ij \in \Emc} d_i d_j.
\end{equation}
This implies for $\{\usf\vsf\} \sim \mathrm{Uni}(\Emc)$ an edge selected uniformly at random $\Ebb[d_{\usf}d_{\vsf}] = \frac{1}{|\Emc|} \sum_{ij \in \Emc} d_i d_j$. It follows that the assortativity \cite{New2002} is, for $\wsf \sim \mathrm{Uni}(\Vmc)$ a vertex selected uniformly at random, 
\begin{equation}
\alpha(G) \equiv \mathrm{Corr}(d_{\usf},d_{\vsf})
= \frac{s(G)/|\Emc| - \Ebb[d_{\wsf}]^2}{\mathrm{Var}(d_{\wsf})}.
\end{equation}
\end{definition}
It is evident that maximizing the $S$-metric is equivalent to maximizing assortativity over a degree class:
\begin{equation}
\Wmc^{(n)}_{\dbf,\rm opt} \equiv \argmax_{G \in \Wmc^{(n)}_{\dbf}}\left( s(G) \right) = \argmax_{G \in \Wmc^{(n)}_{\dbf}}\left( \alpha(G) \right).
\end{equation}

Here, $\Wmc^{(n)}_{\dbf,\rm opt}$ denotes those graphs achieving maximum assortativity over $\Wmc^{(n)}_{\dbf}$.  If there is a unique such graph it is denoted $G_{\dbf,\rm opt}^{(n)}$.

\subsection{Contributions and outline}
The rest of the paper is organized as follows. \secref{rewiring} studies  several greedy rewiring heuristics, each with the goal of identifying a graph of maximum assortativity over the degree class. Counterexamples are presented showing each of the rewiring heuristics may fail to identify such a graph. \secref{wiring} examines the greedy wiring heuristic of Li and Alderson \cite{LiAld2005} designed to identify a graph of maximum assortativity over the degree class. We present a counterexample showing the heuristic may fail to produce a graph in the degree class, and also present a counterexample showing that the heuristic may produce a graph in the class that is not maximally assortative. Both \secref{rewiring} and \secref{wiring} present tabulations of the number of counterexamples of the various types for graphs of order up to $n=9$. \secref{conclusion} contains concluding remarks.

\section{Rewiring}
\label{sec:rewiring}

For a degree class $\Wmc^{(n)}_{\dbf}$ and an initial graph $G_0 \in \Wmc^{(n)}_{\dbf}$, a rewiring heuristic produces a sequence of graphs $(G_0,\ldots,G_T)$, each graph in $\Wmc^{(n)}_{\dbf}$, where $G_{t+1}$ is obtained from $G_t$ by selecting two edges (connecting four distinct vertices) from $G_t$, say $(ij,kl)$, and forming $G_{t+1}$ with $(ij,kl)$ replaced by either $(ik,jl)$ or $(il,jk)$.  Any rewiring is invalid if the resulting graph is either disconnected or has multiple edges, i.e., not in $\Wmc^{(n)}_{\dbf}$.

\subsection{Greedy rewiring heuristics}
\label{sec:grh}

A stochastic rewiring heuristic involves selecting the two edges $(ij,kl)$ at random. While simple to implement, stochastic rewiring has no guarantee on efficiency. This observation lead to the focus of this paper, greedy rewiring heuristics.  Fix $G \in \Wmc^{(n)}_{\dbf}$ and four distinct vertices $\{i,j,k,l\}$, such that $G$ has edges $(ij,kl)$. Rewire edges $(ij,kl)$ to produce either graph $G'=G(ik,jl)$ or $G'=G(il,jk)$; the arguments denote the two new edges replacing edges $(ij,kl)$.  Rewiring induces a change in the $S$-metric:
\begin{equation}
\label{eq:delta_gg}
\Delta_{G,G'} \equiv s(G') - s(G) = 
\left\{ \begin{array}{ll}
(d_i d_k + d_j d_l) - (d_i d_j + d_k d_l), \; & G'=G(ik,jl) \\
(d_i d_l + d_j d_k) - (d_i d_j + d_k d_l), \; & G'=G(il,jk)
\end{array} \right.
\end{equation}
Given edges $(ij,kl)$, $\Delta_{G,G'}$ in \eqnref{delta_gg} is the scalar difference of the $S$-metric of $G$ and one of the two possible rewirings, $G(ik,jl)$ or $G(il,jk)$, producing distinct $G'$. The greedy rewiring heuristics we introduce below explore both rewirings.

Three greedy rewiring heuristics are developed using $\Delta_{G,G'}$; each yields a neighborhood $\Nmc_{\dbf,G}^{(H)}$ of graphs in a meta-graph on $\Wmc^{(n)}_{\dbf}$ (defined below), where each $G'\in \Nmc_{\dbf,G}^{(H)}$ is achieved by a heuristic $H$ approved single rewiring of $G$.

\begin{itemize}[leftmargin=*]
\itemsep=-2pt
\item {\em A:} Improve (or maintain) $s(G)$: $\Nmc^{(A)}_{\dbf,G}$ holds all simple connected graphs $G'$ obtainable by a single rewiring of $G$ such that $\Delta_{G,G'} \geq 0$. \label{heuristic_a}
\item {\em B:} Maximize $\Delta$: $\Nmc^{(B)}_{\dbf,G}$ holds all simple connected graphs $G'$ obtainable by a single rewiring of $G$ such that $\Delta_{G,G'}$ is maximum over all $G'$.
\item {\em C:} Improve and maximize: $\Nmc^{(C)}_{\dbf,G}$ holds all simple connected graphs $G'$ obtainable by a single rewiring of $G$ such that $\Delta_{G,G'} \geq 0$ and $\Delta_{G,G'}$ is maximum over all $G'$. 
\end{itemize}

\subsection{Meta-graphs for a degree class}
\label{ch6_mg_dc}

Meta-graphs are graphs with vertices corresponding to the (simple and connected) non-isomorphic graphs in a degree class $\Wmc^{(n)}_{\dbf}$, for a given degree sequence $\dbf \in \Dscr^{(n)}$. Taylor \cite{Tay1981} defined the {\em undirected} meta-graph $\hat{\Gmc}^{(n)}_{\dbf} = (\Wmc^{(n)}_{\dbf},\hat{\Emc}_{\dbf})$, where edges are added between all pairs of graphs related by edge rewiring, i.e., $\{G,G'\} \in \hat{\Emc}_{\dbf}$ iff $G'=G(ik,jl)$ or $G'=G(il,jk)$ for some pair of edges $(ij,kl)$. Taylor proved (Thm.\ 3.3) that $\hat{\Gmc}^{(n)}_{\dbf}$ is connected.  Thus, any graph in $\Wmc^{(n)}_{\dbf}$ is obtainable, starting from any other graph in $\Wmc^{(n)}_{\dbf}$, through a sequence of rewirings, where each graph in the sequence is simple and connected. Note $\hat{\Gmc}^{(n)}_{\dbf}$ may have self-loops as rewiring $G$ may yield $G'$ isomorphic to $G$.

Rewiring heuristics A, B, and C each correspond to {\em directed} meta-graphs.  First, label each graph $G \in \Wmc^{(n)}_{\dbf}$ with its assortativity $\alpha(G)$ (alternately, $s(G)$).  Next, for each heuristic $H \in \{A,B,C\}$, form the directed meta-graph $\hat{\Gmc}_{\dbf,H}^{(n)} \equiv (\Wmc^{(n)}_{\dbf},\hat{\Emc}_{\dbf,H})$, where $\hat{\Emc}_{\dbf,H} \equiv \{ (G,G') \in \hat{\Emc}_{\dbf} | G' \in \Nmc_{\dbf,G}^{(H)} \}$. That is, each rewiring heuristic is represented by retaining (and orienting) the subset of edges in Taylor's meta-graph $\hat{\Gmc}^{(n)}_{\dbf}$ that satisfy the  heuristic.

\begin{figure}[H]
\centering
\includegraphics[width=0.15\textwidth]{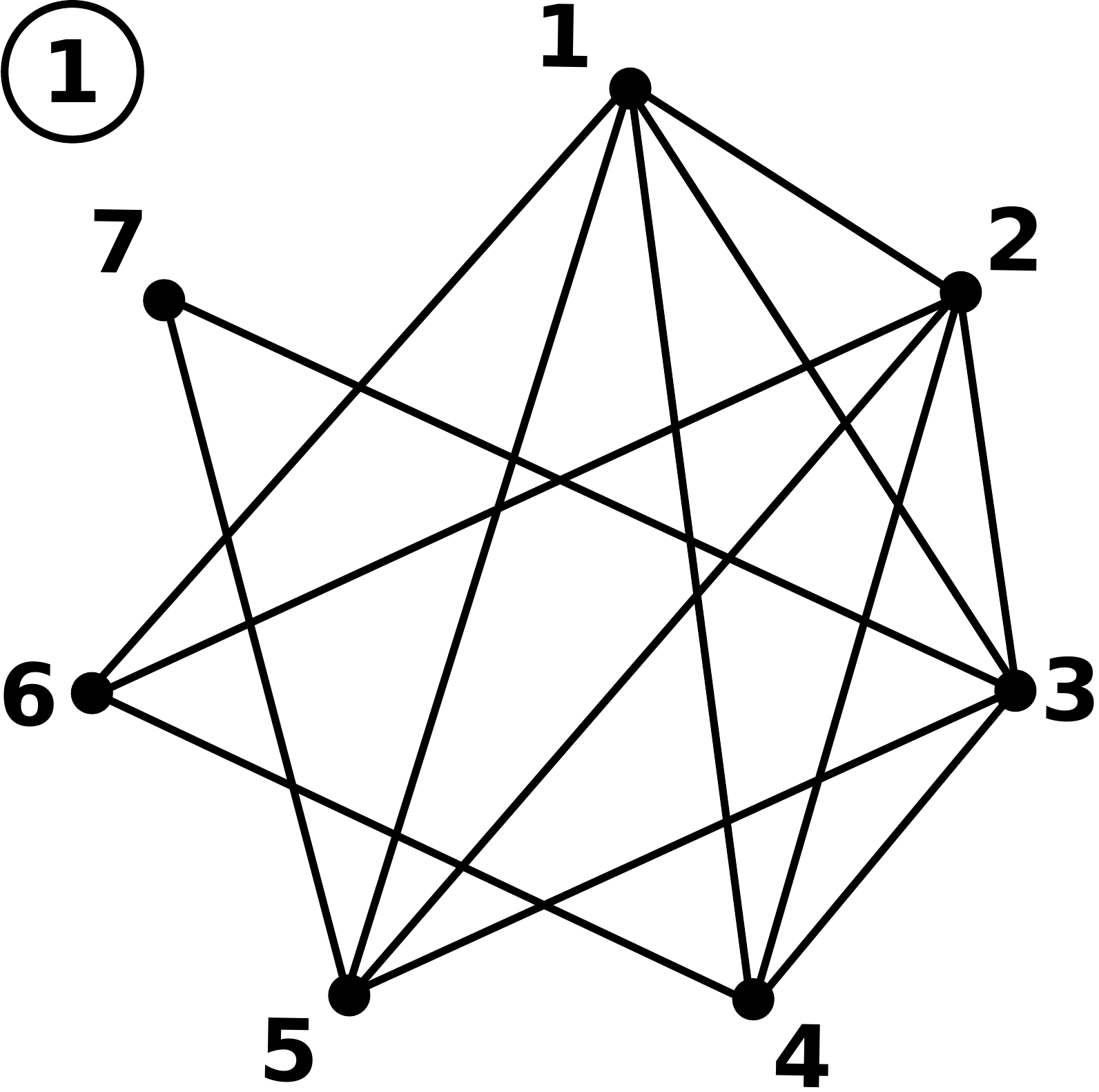}
\hspace{2em}
\includegraphics[width=0.15\textwidth]{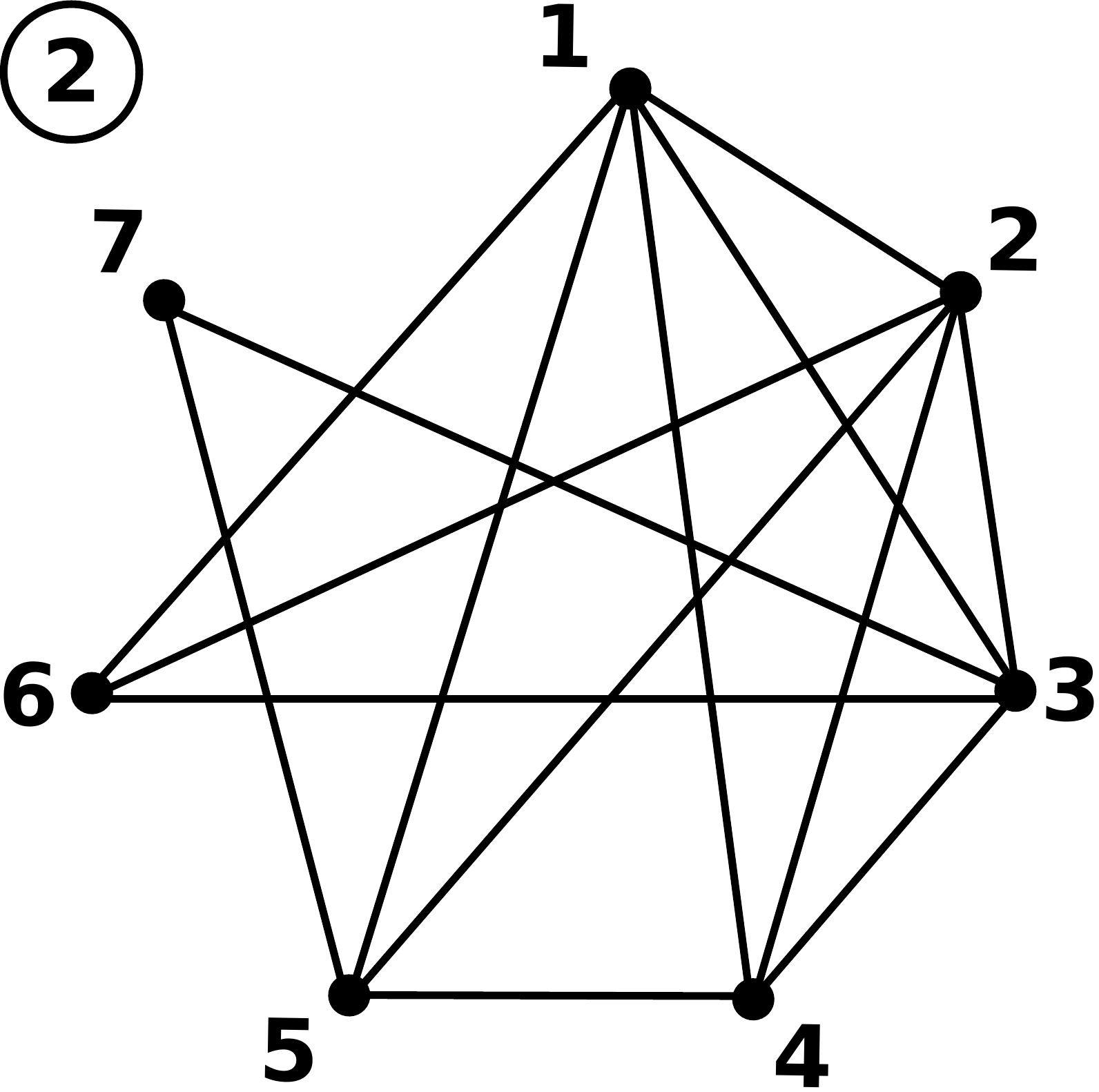}
\hspace{2em}
\includegraphics[width=0.15\textwidth]{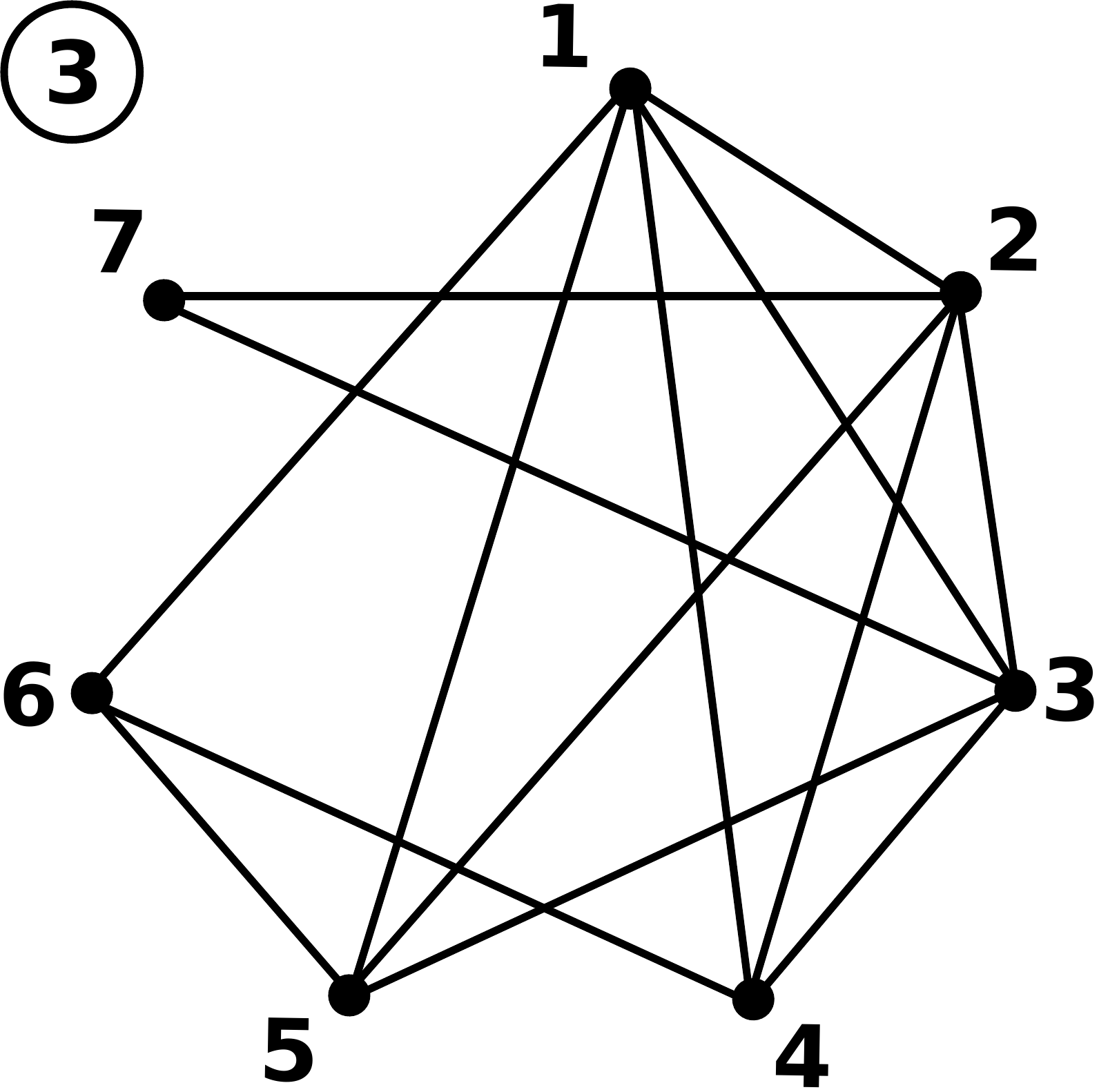}
\hspace{2em}
\includegraphics[width=0.15\textwidth]{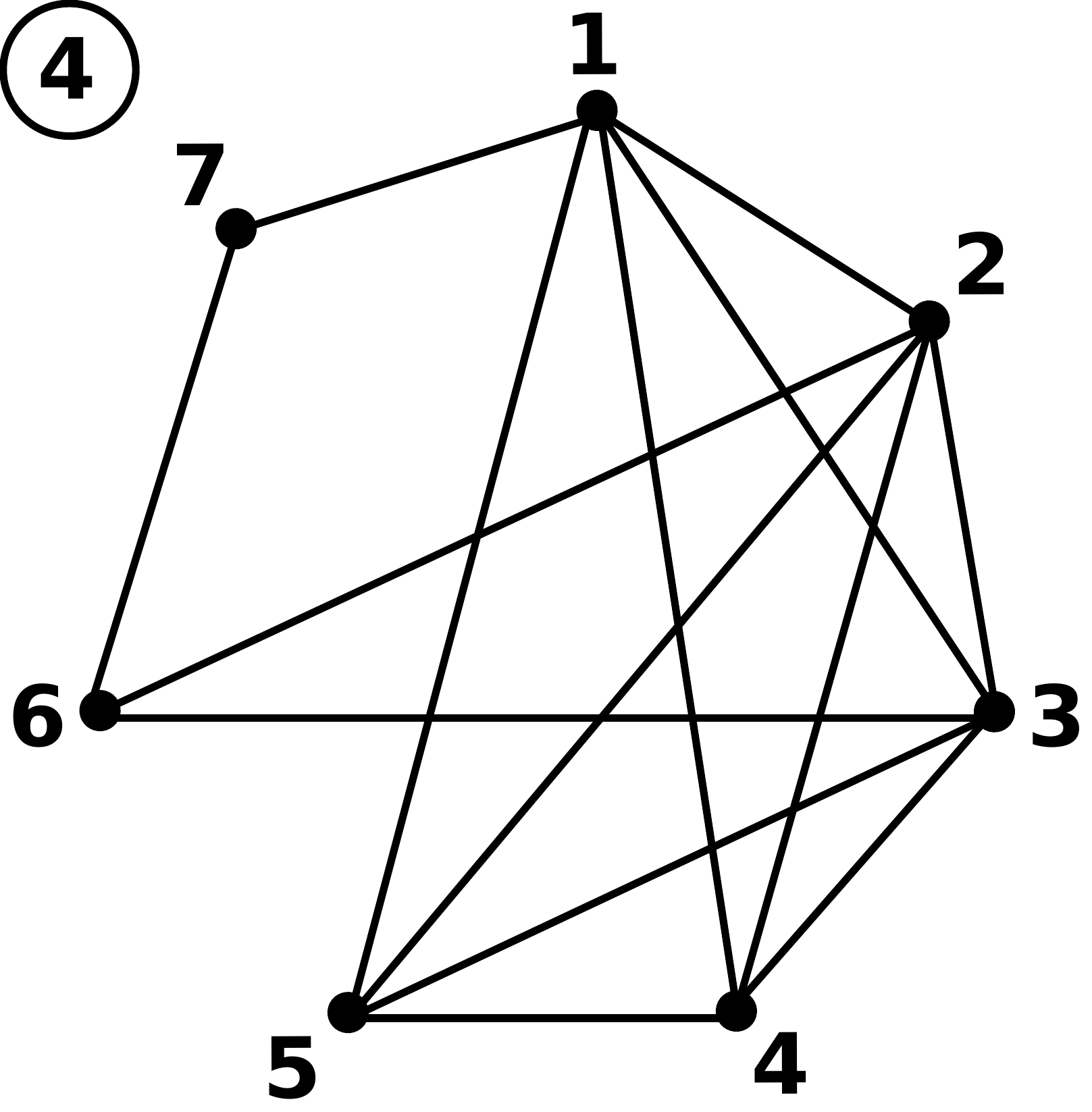}
\hspace{2em}
\includegraphics[width=0.15\textwidth]{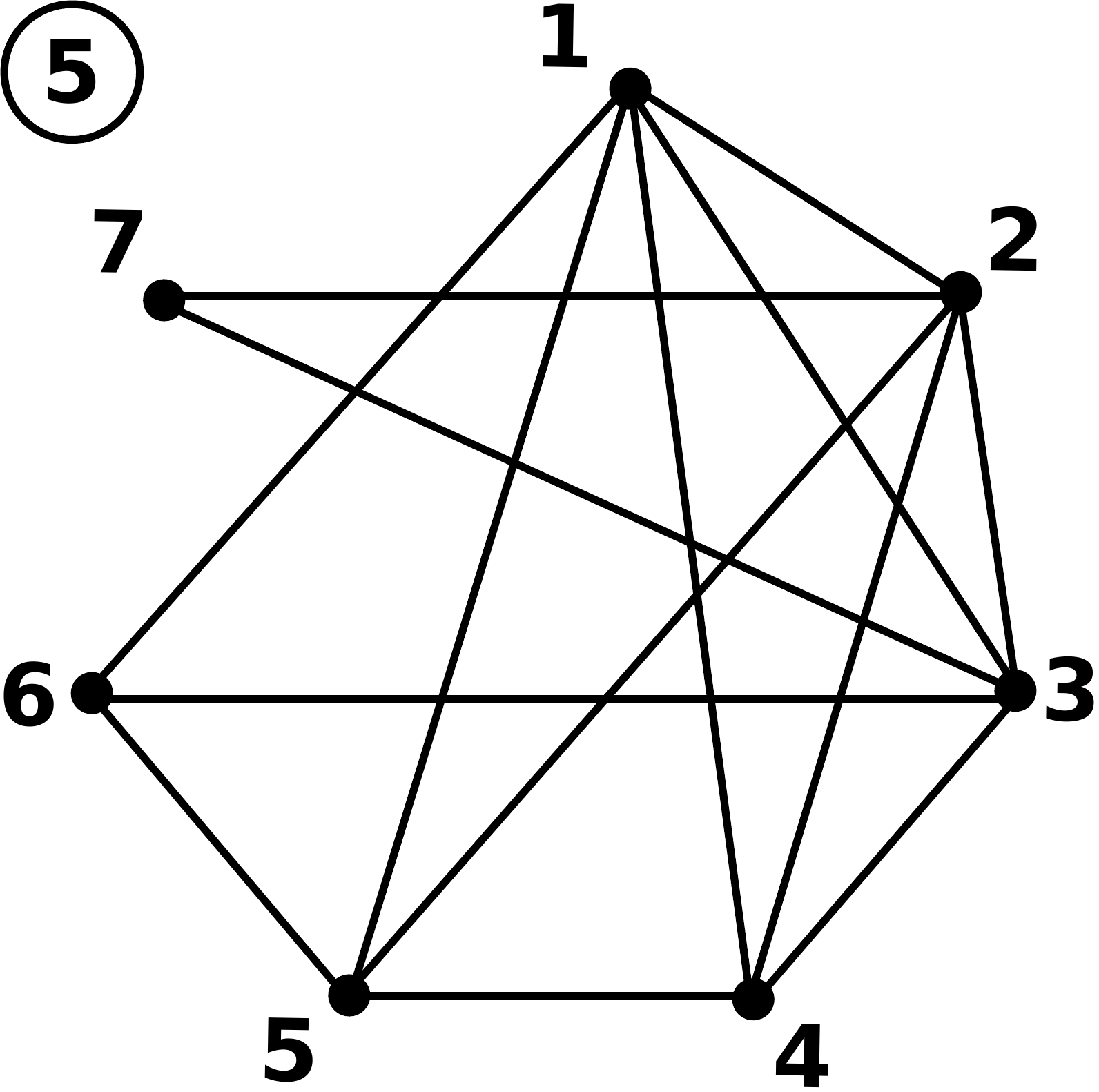}
\hspace{2em}
\includegraphics[width=0.15\textwidth]{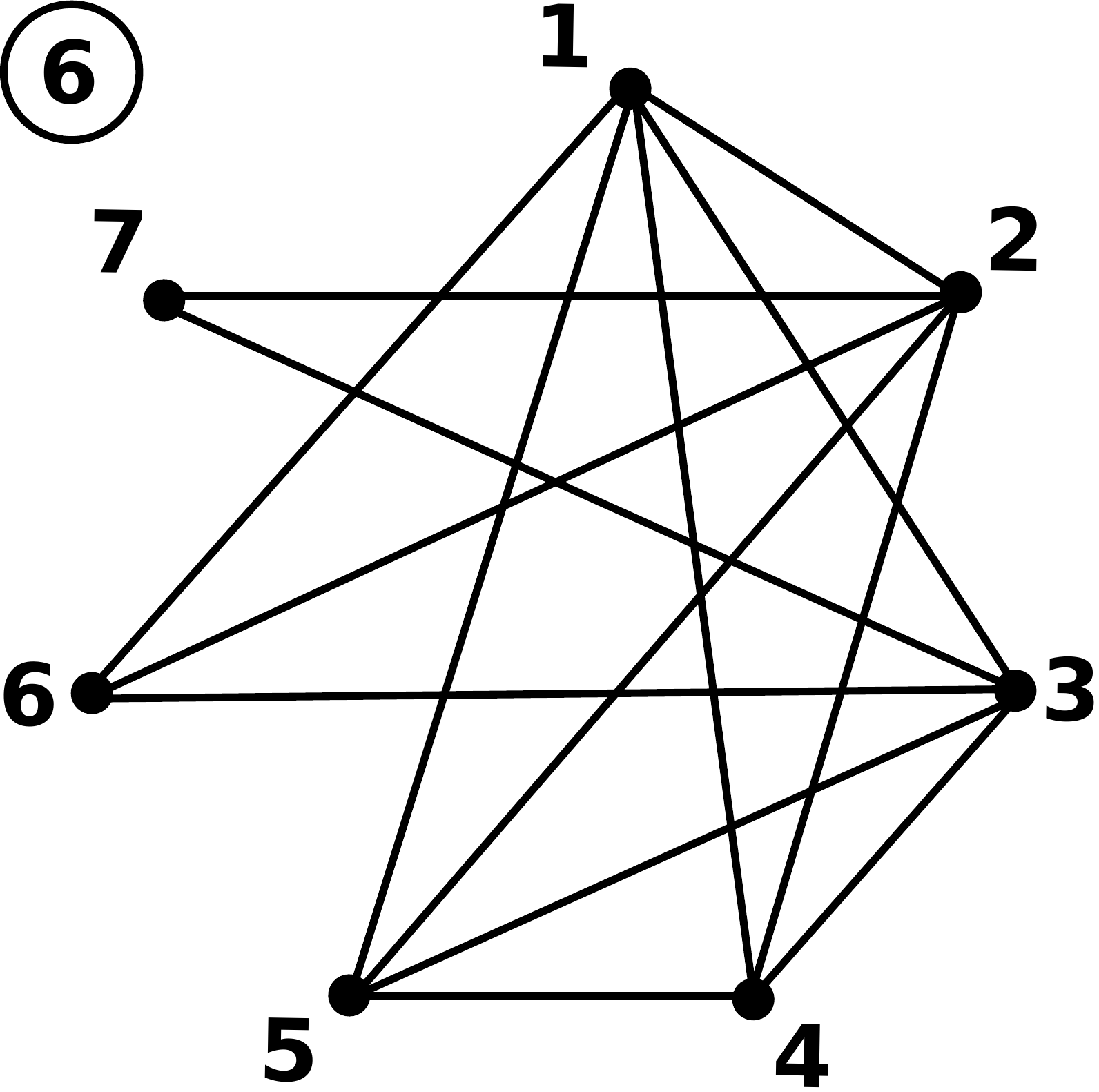}
\hspace{2em}
\includegraphics[width=0.15\textwidth]{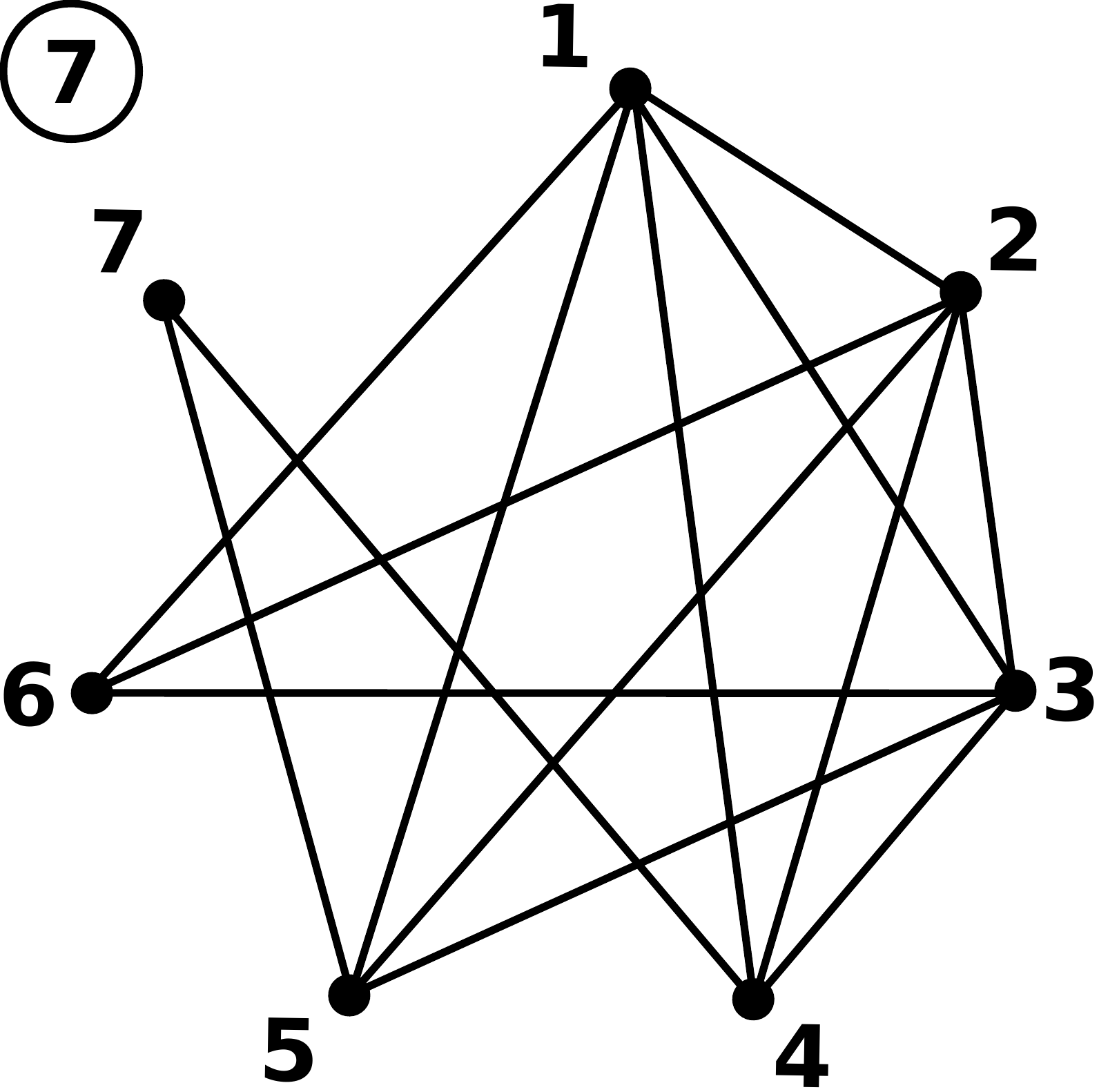}
\caption{From top left to bottom right the graphs corresponding to vertices $1$, $2$, $3$, $4$, $5$, $6$, and $7$ in $\hat{\Gmc}^{(7)}_{\dbf}$, see \figref{mg_Amg_Bmg_Cmg_5554432_1}.}
\label{fig:G_5554432}
\end{figure}

\begin{figure}[H]
\centering
\includegraphics[width=0.15\textwidth]{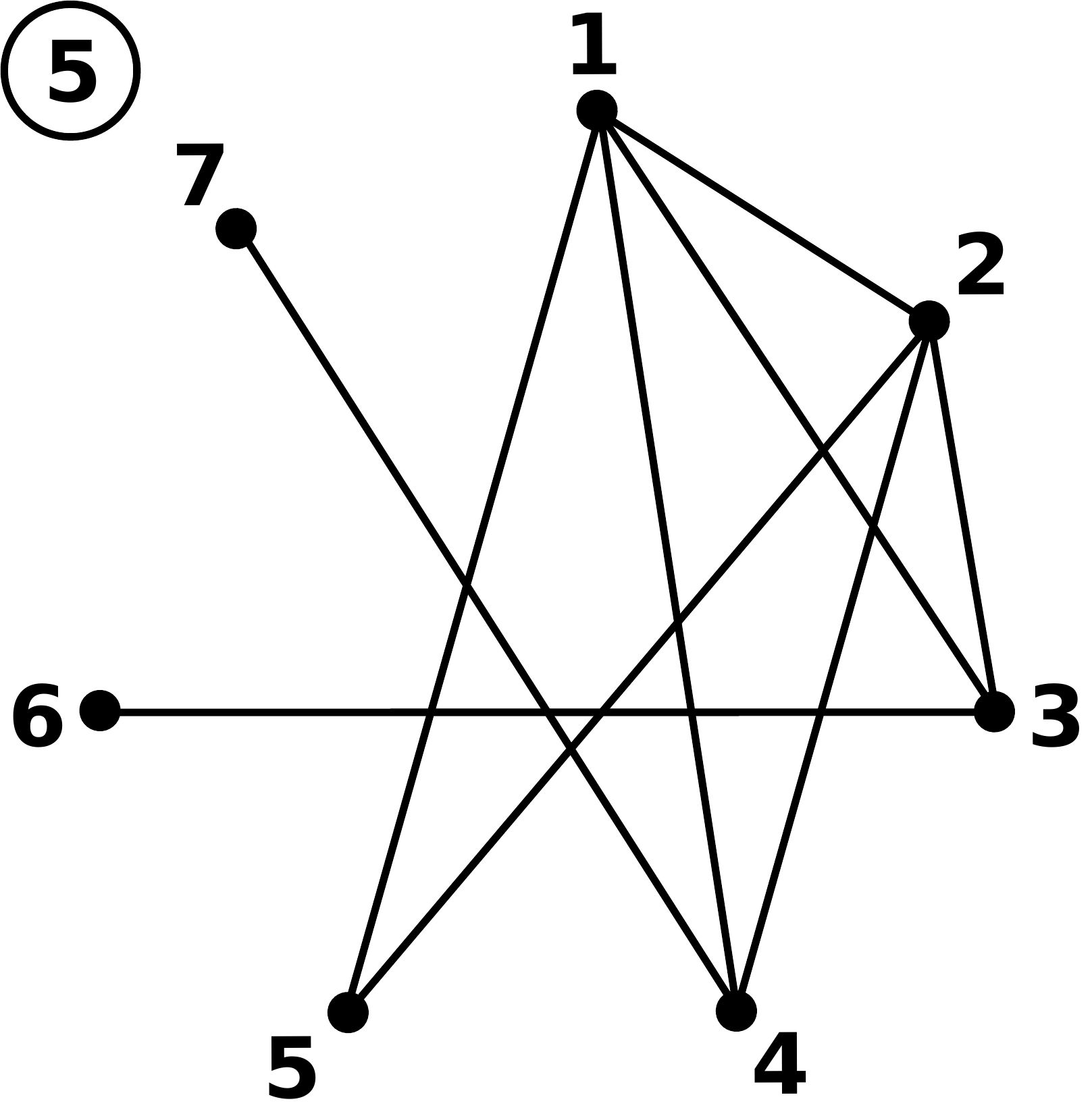}
\hspace{2em}
\includegraphics[width=0.15\textwidth]{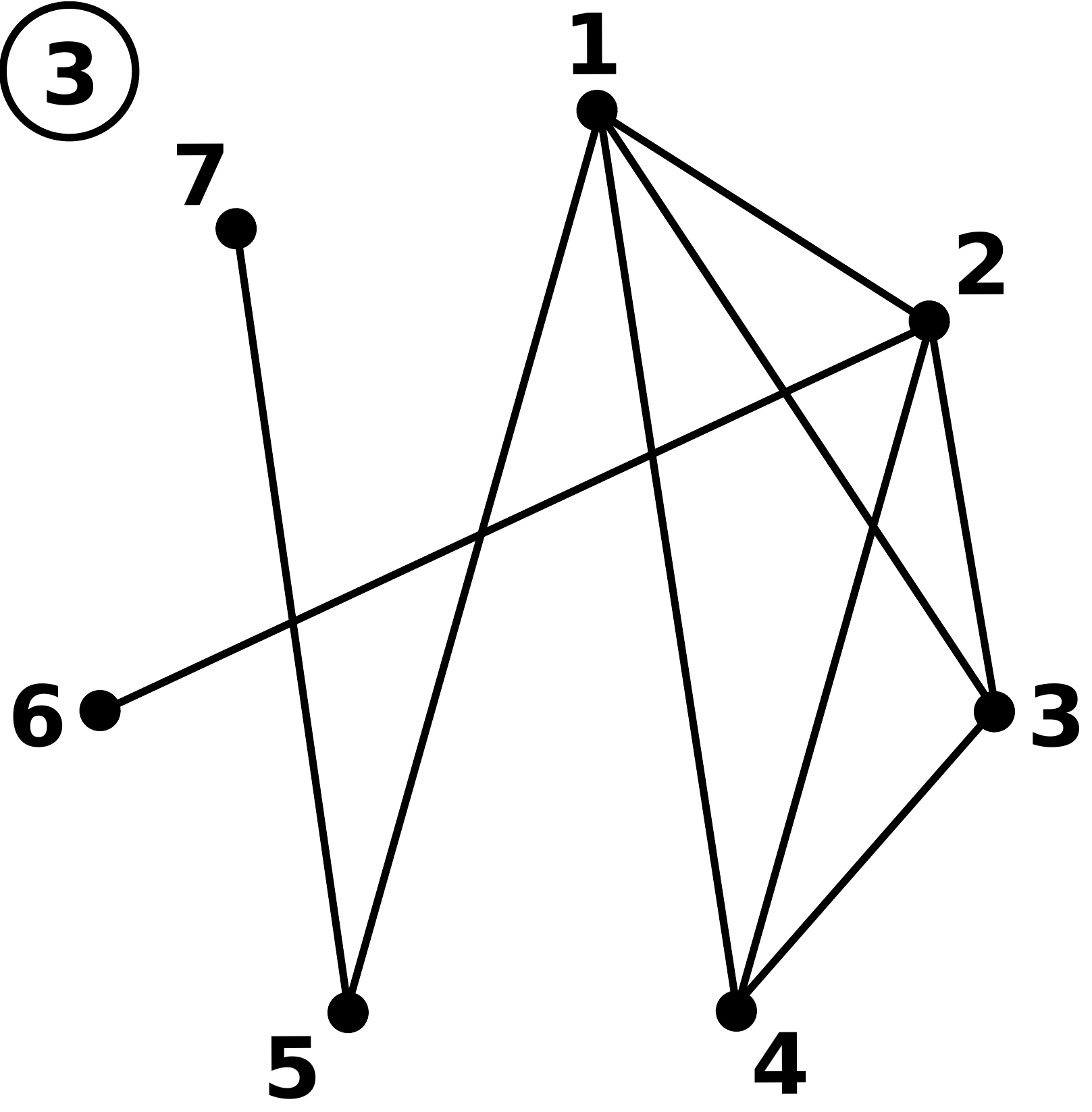}
\caption{ 
From left to right: 
$i)$ initial graph $G_{0,B}$, node $5$ in $\hat{\Gmc}^{(7)}_{\acute{\dbf}}$ $ii)$ target graph $G_{\acute{\dbf},\rm opt}^{(7)}$, node $3$ in $\hat{\Gmc}^{(7)}_{\acute{\dbf}}$, see \figref{mg_Amg_Bmg_Cmg_5554432_1}.}
\label{fig:G_4433211}
\end{figure}

\begin{figure}[H]
\centering
\includegraphics[width=1.0\textwidth]{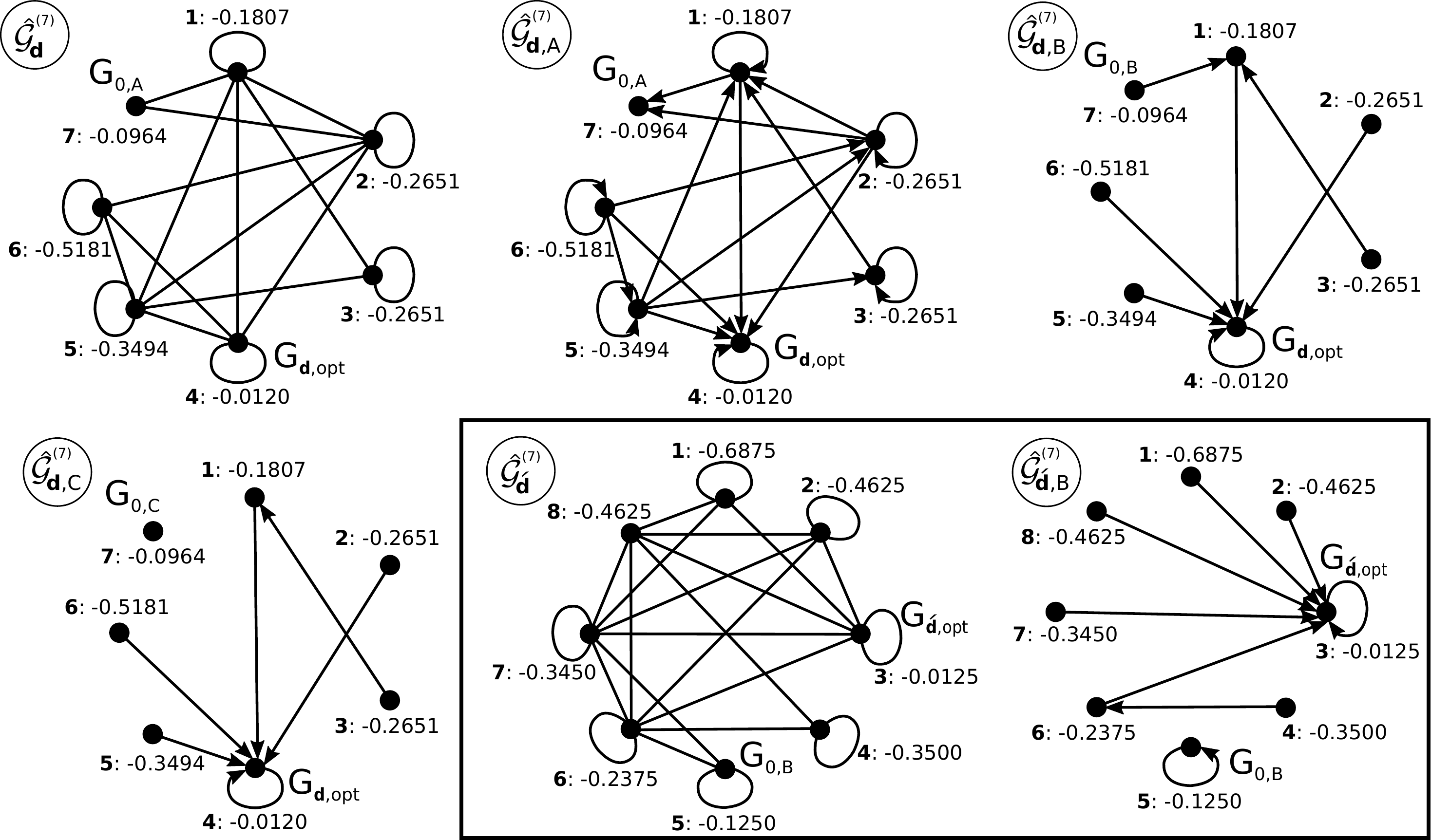}
\caption{The meta-graphs above are, from top left to bottom right: 
$i)$ $\hat{\Gmc}^{(7)}_{\dbf}$, 
$ii)$ $\hat{\Gmc}^{(7)}_{\dbf,A}$, 
$iii)$ $\hat{\Gmc}^{(7)}_{\dbf,B}$,
$iv)$ $\hat{\Gmc}^{(7)}_{\dbf,C}$ with $\dbf = (5,5,5,4,4,3,2)$, 
$v)$ $\hat{\Gmc}^{(7)}_{\acute{\dbf}}$,
$vi)$ $\hat{\Gmc}^{(7)}_{\acute{\dbf},B}$ with $\acute{\dbf}=(4,4,3,3,2,1,1)$.
The number to the right of each vertex id is the assortativity of the corresponding graph.}
\label{fig:mg_Amg_Bmg_Cmg_5554432_1}
\end{figure}

\subsection{Rewiring heuristic counterexamples}
One might hope that (one or more of) the rewiring heuristics would provide a guarantee that, for any initial graph $G_0 \in \Wmc^{(n)}_{\dbf}$, there exists a directed path, following the heuristic, from $G_0$ to one or more graphs in $\Wmc^{(n)}_{\dbf,\rm opt}$. Unfortunately, all three heuristics can fail to achieve this goal, as shown by the counterexamples below. A counterexample for heuristic $H \in \{A,B,C\}$ identifies a $(n,\dbf,G_0)$ triple, with $n \in \Nbb$, $\dbf \in \Dscr^{(n)}$, and $G_0 \in \Wmc^{(n)}_{\dbf}$, such that there is no path from $G_0$ to any graph in $\Wmc^{(n)}_{\dbf,\rm opt}$ in the directed meta-graph $\hat{\Gmc}_{\dbf,H}^{(n)}$.  

Note that these heuristics do not specify a particular rewiring, i.e., each heuristic identifies, in general, a collection of possible neighborhood graphs $\Nmc_G^{(H)}$, where each graph in the neighborhood is consistent with the heuristic.  Thus, a counterexample for the heuristic has the property that the heuristic $H$ would fail to achieve the target set for {\em any} possible choice of $G' \in \Nmc_G^{(H)}$, for each $G$ ``reachable'' from $G_0$.  

\begin{counter_ex}
Fix order $n=7$, degree sequence $\dbf=(5,5,5,4,4,3,2)$, and initial graph $G_{0,A} \in \Vmc^{(7)}_{\dbf}$ (graph $7$ in \figref{G_5554432}). The (unique) graph with maximum assortativity, $G_{\dbf,\rm opt}^{(7)}$, is graph $4$ in \figref{G_5554432}. The meta-graph $\hat{\Gmc}^{(7)}_{\dbf}$ and directed meta-graph under heuristic A, $\hat{\Gmc}^{(7)}_{\dbf,A}$ are graph $1$ and $2$ in \figref{mg_Amg_Bmg_Cmg_5554432_1}. There is no path from $G_{0,A}$ to $G_{\dbf,\rm opt}^{(7)}$ in $\hat{\Gmc}^{(7)}_{\dbf,A}$, and hence no path in $\hat{\Gmc}^{(7)}_{\dbf,C}$ (graph $4$ in \figref{mg_Amg_Bmg_Cmg_5554432_1}).  Thus, $(n,\dbf,G_{0,A})$ is a counterexample for heuristics $A$ and $C$. 
\label{ce:ce_rewire_1}
\end{counter_ex}

This counterexample asserts that graph $G_{0,A}$ has locally (i.e., over graphs adjacent to $G_{0,A}$ in $\hat{\Gmc}^{(7)}_{\dbf,A}$) maximal but not globally (i.e., over $\Wmc^{(7)}_{\dbf}$) maximum assortativity. To see that $G_{0,A}$ is locally maximal, \tabref{mgA_mgB} lists possible pairs of edges from $G_{0,A}$ which if rewired as $G'=G_{0,A}(ik,jl)$ or $G'=G_{0,A}(il,jk)$ maintain graph simplicity and connectivity: $\Delta_{G_{0,A},G'} < 0$ for each possible $G'$. 
\begin{table}[H]
\begin{center}
\footnotesize{
  \begin{tabular}{c|cc|cc}
    \hline
    $(ij,kl)$ & $(ik,jl)$ & $\Delta_{G_{0,A},G'}$ & $(il,jk)$ & $\Delta_{G_{0,A},G'}$ \\ \hline
    $(43,57)$ & $(45,37)$ & $-2$ & $(47,35)$ &*\\
    $(42,57)$ & $(45,27)$ & $-2$ & $(47,25)$ &*\\
    $(41,57)$ & $(45,17)$ & $-2$ & $(47,15)$ &*\\

    $(47,53)$ & $(45,73)$ & $-2$ & $(43,75)$ &*\\
    $(47,52)$ & $(45,72)$ & $-2$ & $(42,75)$ &*\\
    $(47,51)$ & $(45,71)$ & $-2$ & $(41,75)$ &*\\

    $(47,63)$ & $(46,73)$ & $\bm{-1}$ & $(43,67)$ &*\\
    $(47,62)$ & $(46,72)$ & $\bm{-1}$ & $(42,76)$ &*\\
    $(47,61)$ & $(46,71)$ & $\bm{-1}$ & $(41,76)$ &*\\

    $(57,63)$ & $(56,73)$ & $\bm{-1}$ & $(53,67)$ &*\\
    $(57,62)$ & $(56,72)$ & $\bm{-1}$ & $(52,76)$ &*\\
    $(57,61)$ & $(56,71)$ & $\bm{-1}$ & $(51,76)$ &*\\
    \hline
\end{tabular}
\caption{Rewirings of edge pairs $(ij,kl)$ (left) of $G_{0,A}$, along with $\Delta_{G_{0,A},G'}$ for $G'=G_{0,A}(ik,jl)$ (middle) or $G'=G_{0,A}(il,jk)$ (right). Bold entries maximize $\Delta_{G_{0,A},G'}$, * indicates rewirings which violate graph simplicity or connectivity.}
\label{tab:mgA_mgB}
}
\end{center}
\end{table}

\begin{counter_ex}
Fix order $n=7$, degree sequence $\acute{\dbf}=(4,4,3,3,2,1,1)$, and initial graph $G_{0,B} \in \Wmc^{(7)}_{\acute{\dbf}}$ (graph $1$ in \figref{G_4433211}). The (unique) graph with maximum assortativity, $G_{\acute{\dbf},\rm opt}^{(7)}$, is graph $2$ in \figref{G_4433211}. The meta-graph $\hat{\Gmc}^{(7)}_{\acute{\dbf}}$ and directed meta-graph under heuristic B, $\hat{\Gmc}^{(7)}_{\acute{\dbf},B}$, are graph $5$ and $6$ in \figref{mg_Amg_Bmg_Cmg_5554432_1}. There is no path from $G_{0,B}$ to $G_{\acute{\dbf},\rm opt}^{(7)}$ in $\hat{\Gmc}^{(7)}_{\acute{\dbf},B}$.  Thus, $(n,\acute{\dbf},G_{0,B})$ is a counterexample for heuristic $B$. 
\label{ce:ce_rewire_2}
\end{counter_ex}

This counterexample asserts that graph $G_{0,B}$ has locally maximal but not globally maximum assortativity.  This can be seen by enumerating all possible pairs of edges from $G_{0,B}$ which if rewired maintain graph simplicity and connectivity, and showing the (unique) optimal choice to maximize $\Delta_{G,G'}$ produces a new graph $G'$ isomorphic to $G_{0,B}$; this enumeration is omitted due to space constraints. The isomorphism between $G_{0,B}$ and $G'$ produces the self-loop in $\hat{\Gmc}^{(7)}_{\acute{\dbf},B}$ at the vertex corresponding to $G_{0,B}$. 

Counterexamples \ceref{ce_rewire_1} and \ceref{ce_rewire_2} were found via exhaustive search. We enumerated the class of non-isomorphic simple connected graphs of $n$ vertices and degree sequence $\dbf$, $\Wmc^{(n)}_{\dbf}$, using the tool {\sf geng} \cite{McKPip2014}. Letting $G_i$ denote the $i_{th}$ graph in $\Wmc^{(n)}_{\dbf}$ corresponding to the $i_{th}$ vertex in meta-graph $\hat{\Gmc}_{\dbf,H}^{(n)}$, under greedy rewiring heuristic $H$, we enumerate all non-isomorphic edge rewirings of $G_i$. For each rewiring $G'_i$ of $G_i$ that satisfies heuristic $H$ we check for isomorphism of $G'_i$ with $G_k\in\Wmc_{\dbf}^{(n)}$. If $G'_i$ is isomorphic with $G_k$, we add directed edge $(i,k)$ to meta-graph $\hat{\Gmc}_{\dbf,H}^{(n)}$. Upon completing this procedure for all $G\in\Wmc_{\dbf}^{(n)}$, we check that a path exists from each vertex in $\hat{\Gmc}_{\dbf,H}^{(n)}$ to a vertex in $\Wmc^{(n)}_{\dbf,\rm opt}$. Using this procedure we generated \tabref{ce_counts}, which counts the number of counterexamples for each of the greedy rewiring heuristics.
\begin{table}[h]
\begin{center}
\begin{tabular}{r|rr|rr|rr|rr}
\hline
& \multicolumn{2}{c}{overall} & \multicolumn{2}{c}{heuristic A} & \multicolumn{2}{c}{heuristic B} & \multicolumn{2}{c}{heuristic C} \\ \hline
$n$ & $|\Wmc^{(n)}|$ & $|\mathscr{D}^{(n)}|$ & $\#G$ & $\#\dbf$ & $\#G$ & $\#\dbf$ & $\#G$ & $\#\dbf$ \\ \hline
$6$ & $112$ & $68$ & $0$ & $0$ & $0$ & $0$ & $0$ & $0$\\
$7$ & $853$ & $236$ & $2$ & $2$ & $1$ & $1$ & $2$ & $2$\\
$8$ & $11,117$ & $863$ & $13$ & $12$ & $15$ & $8$ & $20$ & $12$\\
$9$ & $261,080$ & $3,137$ & $149$ & $80$ & $1045$ & $67$ & $1100$ & $80$\\
\hline
\end{tabular}
\caption{Rewiring heuristic counterexample counts: The number of distinct graphs $|\Wmc^{(n)}|$ and degree sequences $|\mathscr{D}^{(n)}|$, followed by the number of distinct graphs ($\#G$) and degree sequences ($\#\dbf$) that are counterexamples for heuristics A, B, C, for $n\in\{6,7,8,9\}$.}
\label{tab:ce_counts}
\end{center}
\end{table}

\section{Wiring}
\label{sec:wiring}

If a degree sequence $\dbf$ satisfies the Erd\H{o}s Gallai theorem there exists one or more simple connected graphs with that degree sequence, i.e., $\Wmc^{(n)}_{\dbf} \neq \emptyset$ \cite{ErdGal1960}.  Given such a $\dbf$, a wiring heuristic produces a sequence of graphs $(\tilde{G}_0,\ldots,\tilde{G}_T)$, with $\tilde{G}_0$ the empty graph, such that $\tilde{G}_{t+1}$ is formed from $\tilde{G}_t$ by adding one edge, subject to the constraint that no vertex $j \in \Vmc$ is ever assigned a degree exceeding its target $d_j$. It is typical to consider each vertex $j$ in graph $\tilde{G}_t$ as having $d_j$ ``stubs'' of which some number $\tilde{d}_j$ hold edges, and the remainder, $\delta_j \equiv d_j - \tilde{d}_j$, are available for wiring. The goal of the wiring heuristics is to obtain a graph of maximum assortativity, i.e., $\tilde{G}_T \in \Wmc^{(n)}_{\dbf,\rm opt}$ given $\dbf$.

\subsection{Greedy wiring heuristic}
Li and Alderson \cite{LiAld2005} developed the elegant greedy wiring heuristic in \algoref{s_max} which, given a degree sequence $\dbf$ is intended to produce a graph $\tilde{G}$ that is $i)$ feasible, i.e., that is in $\Wmc^{(n)}_{\dbf}$, and $ii)$ optimal, i.e., in $\Wmc^{(n)}_{\dbf,\rm opt}$. Although the heuristic performs well on most inputs, the following section will present counterexamples demonstrating neither property is guaranteed for all $\dbf$.

Each potential edge, hereafter a ``pedge'', is denoted by the ordered pair $(ij)$ with $i < j$.  The basic idea is to select from set of all pedges $\Omc$ those with the largest endpoint degree product, $\Mmc$ (\linref{mmc}), after removing from $\Omc$ and $\Mmc$ pedges in $\Mmc$ without available (unwired) stubs $\Fmc$ (\linref{fmc}).  If pedges remain then further ties are broken by first (then second) selecting the pedge $(ij)$ with the most unwired stubs $\delta_i$ ($\delta_j$).  Vertices $[n]^+$ are partitioned into $\Rmc,\Qmc$, where $\Rmc$ ($\Qmc$) holds any vertex with one or more (no) edges.  If the pedge $(ij)$ has $i \in \Rmc$ and $j \in \Qmc$ then the edge is added and vertex $j$ moves from $\Qmc$ to $\Rmc$ (\linref{ABedge}). Else $\Rmc$ holds both $i$ and $j$ and we must check the ``tree condition'', $(d_{\Bmc} \neq  (2|\Bmc|-\delta_{\Rmc}))$, and ``disconnected cluster condition'', $(\delta_{\Rmc} \neq 2)$, in \linref{treedcc}.

The ``tree condition'' is required since at any point in wiring the graph, connecting the vertices in $\Qmc$ to the $\delta_{\Rmc} := \sum_{k\in\Rmc}\delta_k$ free stubs in $\Rmc$ requires $\delta_{\Rmc}$ acyclic graphs and at least $2|\Qmc|-\delta_{\Rmc}$ free stubs in $\Qmc$. As $\delta_{\Rmc}$ decreases, the number of free stubs in $\Qmc$ required to connect the vertices in $\Qmc$ to the free stubs in $\Rmc$ increases. Letting $\delta_{\Qmc} := \sum_{k\in\Qmc}\delta_k$, if an edge is added between vertices $i$ and $j$ in $\Rmc$ which results in $d_{\Qmc} < 2|\Qmc|-\delta_{\Rmc}$ then there are not enough free stubs in $\Qmc$ to connect all the vertices in $\Qmc$ to those in $\Rmc$, entailing that $\tilde{G}$ is disconnected. The ``disconnected cluster condition'' is required since wiring an edge between the only two free stubs in $\Rmc$ entails $\tilde{G}$ is disconnected, as no additional vertices in $\Qmc$ can be attached to those in $\Rmc$ (see \cite{LiAld2005}). Regardless of whether or not the conditions in \linref{treedcc} are satisfied, the pedge $(ij)$ being considered for wiring is removed from the set $\Omc$ in \linref{rm_pedge}.

\begin{algorithm}[H]
\footnotesize{
\caption{Greedy wiring heuristic (adapted from \cite{LiAld2005})} 
\label{alg:s_max}
\algrenewcommand\algorithmicindent{0.7em}
\begin{algorithmic}[1]
\State \textbf{require:} $\dbf = (d_1,\ldots,d_n)$ with $d_1 \geq \cdots \geq d_n$
\State $\Rmc := \{1\}$, $\Qmc := \{2,\ldots,n\}$, $\tilde{\Emc}:=\{\}$, $\tilde{G} := (\Rmc,\tilde{\Emc})$, $\Omc := \{(ij) : 1 \leq i < j \leq n\}$
\While{$\Omc \neq \emptyset$}
	\State $\Mmc := \argmax_{(ij)\in \Omc}(d_i d_j)$ \label{lin:mmc}
	\State $\Fmc := \{ (ij) \in \Mmc : \delta_i \delta_j = 0\}$
	\State $\Omc := \Omc \backslash \Fmc$, $\Mmc := \Mmc \backslash \Fmc$ \label{lin:fmc}
	\If{$\Mmc \neq \emptyset$}
		\State $\Mmc' := \argmax_{(ij) \in \Mmc} \delta_i$
		\label{lin:max_i_stubs}
		\State Select $(ij) \in \argmax_{(ij) \in \Mmc'} \delta_j$
		\label{lin:max_j_stubs}
		\If{$i \in \Rmc$ and $j \in \Qmc$}
			\State $\tilde{\Emc} := \tilde{\Emc} \cup \{ij\}$, $\Rmc := \Rmc \cup \{j\}$, $\Qmc := \Qmc \backslash \{j\}$ \label{lin:ABedge}
		\Else
			\State $d_{\Qmc} := \sum_{k \in \Qmc} d_k$, $\delta_{\Rmc} := \sum_{k \in \Rmc} \delta_k$
			\If {$(d_{\Qmc} \neq  (2|\Qmc|-\delta_{\Rmc})) \wedge (\delta_{\Rmc} \neq 2)$} \label{lin:treedcc}
				\State $\tilde{\Emc} := \tilde{\Emc} \cup \{ij\}$
			\EndIf
	\EndIf
	\State $\Omc := \Omc \backslash \{(ij)\}$ \label{lin:rm_pedge}
	\EndIf
\EndWhile
\State \textbf{return} $\tilde{G}$
\end{algorithmic}
}
\end{algorithm}

\algoref{s_max} is underspecified in \linref{max_j_stubs}, i.e., there may be multiple edges after sorting $\Omc$ by $d_i d_j$, $\delta_i$, and $\delta_j$, and no guidance is provided in \cite{LiAld2005} for selecting a pedge in such a case. To compensate for this, the implementation of \algoref{s_max} in this paper selects {\em all} possible pedge choices, via a breadth first search, returning all possible graphs $\tilde{G}$ that may result from a valid pedge selection in \linref{max_j_stubs}. A degree sequence $\dbf$ is considered to be a counterexample for $i)$ feasibility if none of the returned graphs are in $\Wmc^{(n)}_{\dbf}$, and $ii)$ optimality if at least one returned graph is in $\Wmc^{(n)}_{\dbf}$, yet none are in $\Wmc^{(n)}_{\dbf,\rm opt}$.

\subsection{Wiring heuristic counterexamples}
\label{ds_wire}

\begin{counter_ex}
Fix $n=6$ and $\dbf=(5,4,4,4,4,3)$ (which satisfies the Erd\H{o}s Gallai theorem). The graph $\tilde{G}$ returned by \algoref{s_max} does not have the target degree sequence, i.e., $\dbf_{\tilde{G}} \neq \dbf$, and thus $\tilde{G}$ is not feasible, i.e., $\tilde{G} \not\in \Wmc^{(6)}_{\dbf}$.  
\label{ce:ce_wire_1}
\end{counter_ex}

\begin{figure}[H]
\centering
\includegraphics[width=0.2\textwidth]{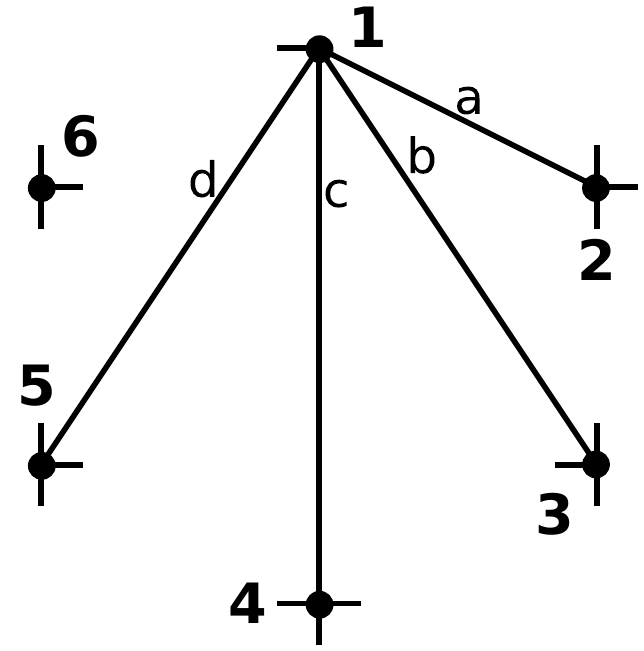}
\includegraphics[width=0.2\textwidth]{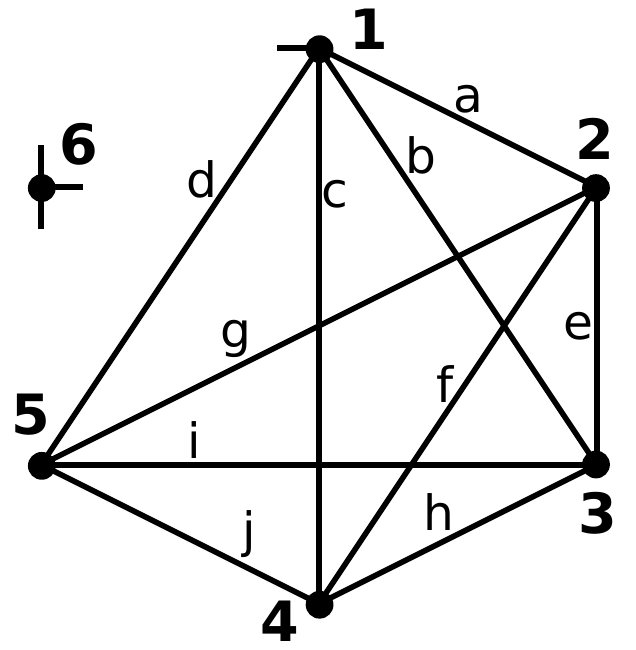}
\includegraphics[width=0.2\textwidth]{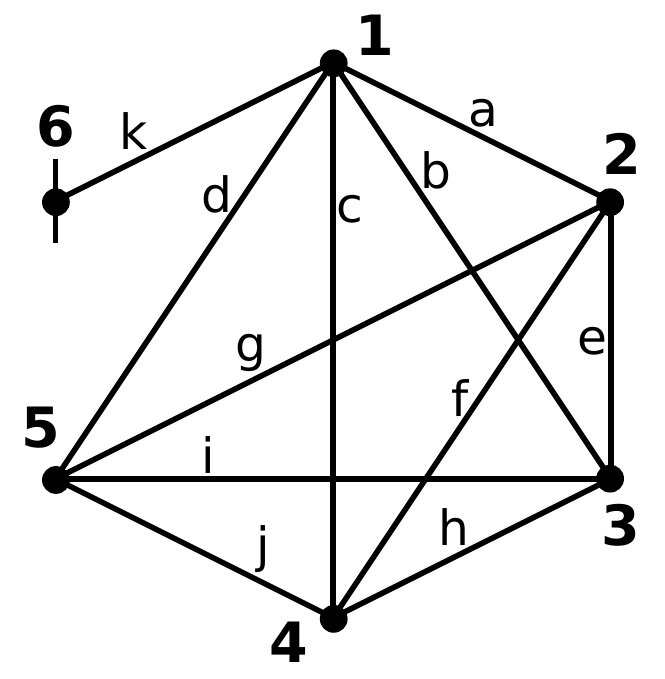}
\caption{Snapshots of the graph wiring in \ceref{ce_wire_1} for $n=6$ and $\dbf=(5,4,4,4,4,3)$ where the edges are added in alphabetical order: From left to right $i)$ $\tilde{G}_4$, $ii)$ $\tilde{G}_{10}$, $iii)$ $\tilde{G}_{11}$.}
\label{fig:ds_ce_G4_G10_G11}
\end{figure}

\begin{proof}
\tabref{ce1_s1_s5_s11} gives the sequence of wirings satisfying $(ij) \in \argmax_{(ij) \in \Omc}(d_i d_j)$, illustrated in \figref{ds_ce_G4_G10_G11}. The first four edges added, namely $(12),\ldots,(15)$, have identical priority as $d_i d_j$, $\delta_i$, and $\delta_j$ are equal for each.  These four may be added in any order without affecting the resulting graph.  The next edges added will be $(23)$, $(24)$, $(25)$, $(34)$, $(35)$, $(45)$.  Finally, $(16)$ will be added, leaving the only two free stubs in the graph on vertex $6$, which can only be wired via a self-loop, thereby violating the requirement that $\tilde{G}$ be simple. 
\end{proof}

\begin{table}[h]
\footnotesize{
\begin{center}
  \begin{tabular}{cccc}
    \hline
    $(ij)$ & $d_i d_j$ & $\delta_i$ & $\delta_j$ \\ \hline
    $(12)$ & $20$ & $5$ & $4$ \\
    $(13)$ & $20$ & $5$ & $4$ \\
    $(14)$ & $20$ & $5$ & $4$ \\
    $(15)$ & $20$ & $5$ & $4$ \\
    \hline
    $(23)$ & $16$ & $3$ & $3$ \\
    $(24)$ & $16$ & $3$ & $3$ \\
    $(25)$ & $16$ & $3$ & $3$ \\
    $(34)$ & $16$ & $3$ & $3$ \\
    $(35)$ & $16$ & $3$ & $3$ \\
    $(45)$ & $16$ & $3$ & $3$ \\
    \hline
    $(16)$ & $15$ & $1$ & $3$ \\
    \hline
  \end{tabular}
\caption{Subset of edge wirings for \ceref{ce_wire_1}. The first set of rows correspond to wirings which are optimal at wiring step $1$. The second set of rows are optimal wirings at wiring step $5$. The final row is the only legal at wiring at step $11$.}
\label{tab:ce1_s1_s5_s11}
\end{center}
}
\end{table}

\begin{counter_ex}
Fix $n=8$ and $\dbf=(6,4,4,4,4,3,2,1)$ (which satisfies the Erd\H{o}s Gallai theorem). The graph $\tilde{G}$ returned by \algoref{s_max} is feasible, but its assortativity is not maximum and thus $\tilde{G}$ is not optimal, i.e., $\tilde{G} \not\in \Wmc^{(8)}_{\dbf,\rm opt}$. 
\label{ce:ce_wire_2}
\end{counter_ex}

\begin{figure*}[h]
\centering
\includegraphics[width=0.2\textwidth]{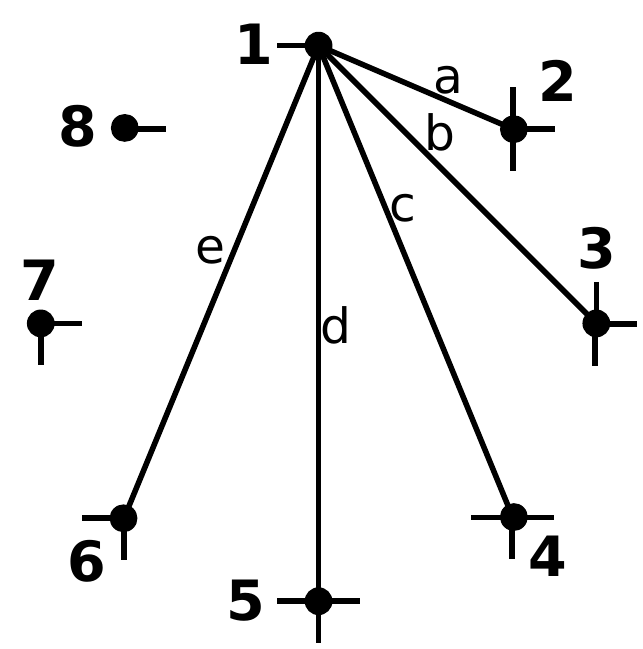}
\includegraphics[width=0.2\textwidth]{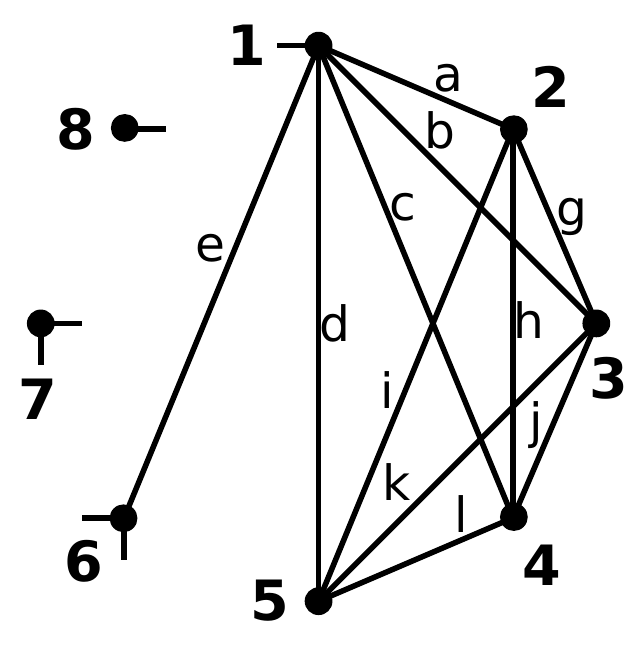}
\includegraphics[width=0.2\textwidth]{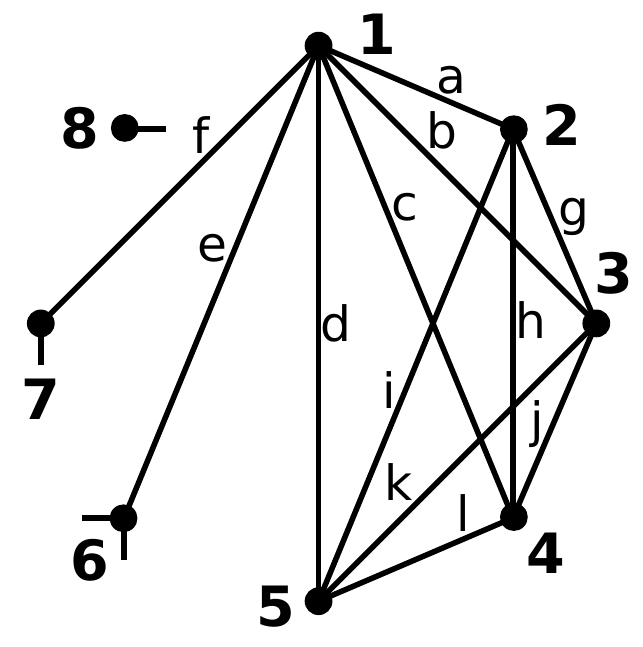}\\
\includegraphics[width=0.2\textwidth]{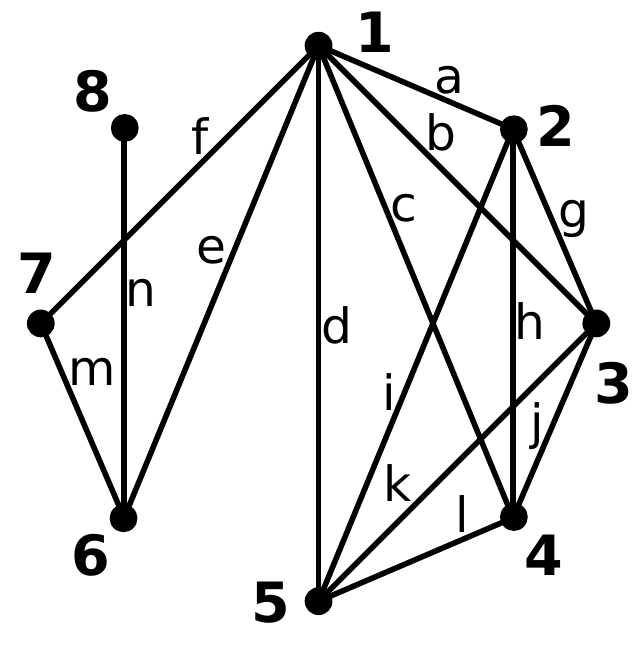}
\includegraphics[width=0.2\textwidth]{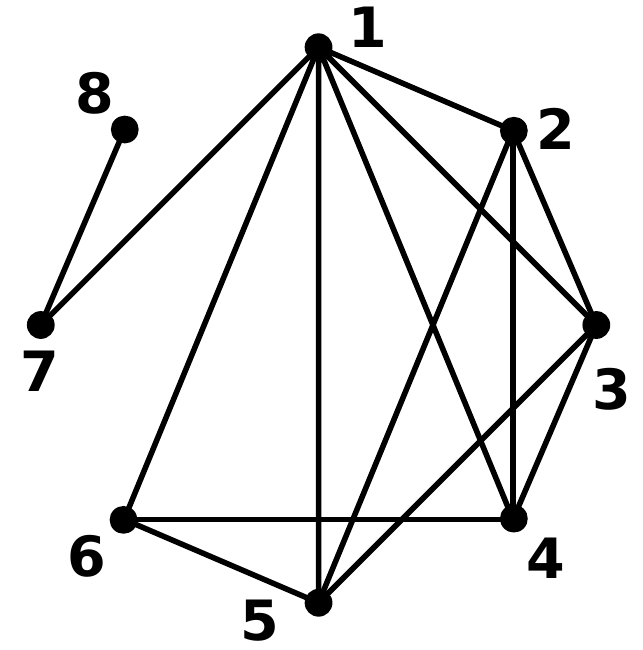}
\caption{Snapshots of the graph wiring in \ceref{ce_wire_2} for $n=8$ and $\dbf=(6,4,4,4,4,3,2,1)$ where the edges are added in alphabetical order: From top left to bottom right $i)$ $\tilde{G}_5$, $ii)$ $\tilde{G}_{11}$, $iii)$ $\tilde{G}_{12}$, $iv)$ $\tilde{G}_{14}$, and $v)$ the maximally assortative graph $G^{(8)}_{\dbf,\rm opt}$.}
\label{fig:alpha_ce_G5_G11_G12_G14_Galpha}
\end{figure*}

\begin{proof}
\label{ch6_ce2_s1_s6_s12_s13}
The proof is similar to \ceref{ce_wire_1}. The partially wired graphs at steps $5$, $11$, $12$, and $14$ are shown in \figref{alpha_ce_G5_G11_G12_G14_Galpha}. The returned graph $\tilde{G} = \tilde{G}_{14}$ achieves the target degree sequence $\dbf$, however its assortativity is not optimal. Namely, $\alpha(\tilde{G}_{14}) =-0.04886$ while $\alpha(G^{(8)}_{\dbf,\rm opt}) = -0.00326$, where $G^{(8)}_{\dbf,\rm opt}$ is graph $5$ in \figref{alpha_ce_G5_G11_G12_G14_Galpha}.
\end{proof}

To find counterexamples \ceref{ce_wire_1} and \ceref{ce_wire_2} we enumerate all degree sequences of order $n$ which satisfy the Erd\H{o}s Gallai theorem. Given a degree sequence $\dbf$ of $n$ vertices, we use \algoref{s_max} to wire target degree sequence $\dbf$. Given the breadth first search, \algoref{s_max} returns a set of graphs $\tilde{\Gmc}$. If $\dbf_{\tilde{G}} \neq \dbf$ for all $\tilde{G}\in\tilde{\Gmc}$, we count $\dbf$ as a feasibility counter example. Otherwise, we check if there exists any $\tilde{G}\in\tilde{\Gmc}$ (obeying $\dbf_{\tilde{G}} = \dbf$) for which $\alpha(\tilde{G}) = \alpha(G_{\dbf,\rm opt}^{(n)})$. If not, we count $\dbf$ as an optimality counterexample. We use this procedure to generate the counts of feasibility and optimality counterexamples in \tabref{wire_ce}.

\begin{table}[h]
\footnotesize{
\begin{center}
\begin{tabular}{rrrr}
\hline
$n$ & $|\mathscr{D}^{(n)}|$ & feasibility & optimality \\ \hline
$5$ & $19$ & $0$ & $0$ \\ 
$6$ & $68$ & $2$ & $0$ \\ 
$7$ & $236$ & $16$ & $0$ \\ 
$8$ & $863$ & $91$ & $4$ \\ 
$9$ & $3,137$ & $443$ & $36$ \\ \hline 
\end{tabular}
\caption{Wiring heuristic counterexample counts: The number of degree sequences $|\mathscr{D}^{(n)}|$, the number of degree sequences for which the returned graph is not feasible, and (if feasible) is not optimal, for $n\in\{5,6,7,8,9\}$.}
\label{tab:wire_ce}
\end{center}
}
\end{table}

\section{Conclusion}
\label{sec:conclusion}
The main point of this paper is to demonstrate the failure of natural greedy heuristics, for both graph rewiring and wiring, to produce connected simple graphs with maximum assortativity over an arbitrary target degree class. Many open questions remain, such as how the relative prevalence of the various classes of counterexamples scale with $n$. One possible direction for future work is to seek to characterize common structural properties of the degree sequences $\dbf \in \mathscr{D}^{(n)}$ comprising the four types of counterexamples given above. 


\bibliographystyle{elsarticle-num}
\bibliography{ipl_bibtex}
\end{document}